# COMPLEMENTARY MEMRISTIVE DEVICE BASED ON A THIN FILM OF MoS$_2$ MONOLAYERS

A. Radoi[1], M. Dragoman[1a], D. Dragoman[2], M. Kusko[1]

[1]*National Research and Development Institute in Microtechnology, Str. Erou Iancu Nicolae*

[2]*Univ. Bucharest, Physics Faculty, P.O. Box MG-11, 077125, Romania*

ABSTRACT

In this manuscript we demonstrate experimentally that a thin film of MoS$_2$ monolayers formed by drop-casting on a gold interdigitated electrode on Si/SiO$_2$ behaves like a complementary memristive device, which is a key device of future crossbar memories. The hysteretic behavior of this device is modulated by light in a spectral range expanding from UV up to IR. In contrast to previous complementary memristive devices based on complicated oxides heterostructures, this device has a simple fabrication procedure and can be controlled by light, in addition to electrical signals.

____________________________________________________________________________c

Corresponding author: mircea.dragoman@imt.ro

ARTICLE

Two-dimensional (2D) transition metal dichalcogenides are a new category of 2D materials discovered after graphene [1]. As in the case of graphene, new physical properties, which are not encountered before, are expected to occur in these materials. Indeed, molybdenum disulphide ($MoS_2$) – the most notorious member of this class of new 2D materials – confirms the expectations due to its amazing properties, not enough exploited or well understood to date. Among them, we note that the bandgap of 2D $MoS_2$ decreases as the number of monolayers increases; monolayer $MoS_2$ is a direct semiconductor, with a bandgap of about 1.9 eV, while its bulk counterpart is an indirect semiconductor with a bandgap of about 1.2 eV. Also, the photodetection properties of 2D $MoS_2$ are impressive, monolayers emitting light with a quantum efficiency $10^4$ times higher compared to the bulk material [2]. Based on the physical properties mentioned above, a wealth of electronic and optoelectronic devices have emerged.

$MoS_2$ monolayers have been first obtained using the same exfoliation procedure as for graphene, and the same back-gate-based electrical configuration was employed to fabricate monolayer or few-monolayer $MoS_2$ transistors [3], photodetectors [4], phototransistors [5], optical memories [6] and even valleytronic devices [7]. However, contrary to field-effect-transistors (FETs) based on graphene, which is a zero-bandgap semiconductor with a relative low number of carriers that obey the Dirac equation, in monolayer $MoS_2$ FETs a saturation region exists in the drain current-drain voltage dependence. This is a result of the existence of a direct-bandgap in monolayer $MoS_2$, an essential feature for nanoscale FET devices which are dedicated especially to digital applications. On the other hand, the rather large number of carriers satisfying the Schrödinger equation monolayer $MoS_2$, together with the presence of many defects and traps, decrease the mobility in $MoS_2$ FETs in comparison to that in graphene.

In this paper, we investigate electrical conduction in thin films containing $MoS_2$ monolayers. Such physical systems are less studied than mechanically exfoliated few- or single-monolayer $MoS_2$, but in the last period the physical properties of these systems have started to receive attention and their first applications, such as mobility engineering [8] and even a FET [9], started to emerge. In what follows, we show that a thin film containing $MoS_2$ monolayers deposited on an Au interdigitated electrode (IDT), which is patterned over a $Si/SiO_2$ substrate behaves as a complementary memristive switch (CMS). In addition, the CMS width of this simple device can be tuned using an optical field directed on the thin film. CMS devices are expected to solve the thorny issue of sneak path problem (see [10], which is mainly a leakage issue) for very-large-scale (VLSI) crossbar memory arrays. Such crossbar arrays are among the most advanced future random access memories (RAMs). Among the first reports on CMSs, we mention the devices based on two antiseries Pt/solid electrolyte/Cu heterostructures [10], the triple layer Pt/TiOx/TiOy/Pt and TiOx/TiON/TiOx/Pt heterostructures based on TiO oxides [11] and the $Ag/GeS_x/Pt$ system [12]. In our case, CMS functionality can be achieved without complicated homo- and heterostructures.

2. METHODS

The $MoS_2$ ultrafine powder was obtained from Graphene Supermaket. An ultrasound dispersion in isopropanol was prepared (0.1 mg•µL$^{-1}$) and 10 µL of it were drop casted onto the IDT, until it reached a thickness of about 10 µm. The IDT gold deposition on the $Si/SiO_2$ substrate was done via sputtering, with the help of the equipment deposition AUTO500BBOC Edwards. The $SiO_2$ substrate with a thickness of 900 nm was deposited by the thermal oxidation method, the Si layer with a thickness of 450 µm having a resistivity of 10 Ω cm. The IDT deposited on $SiO_2$ has 100 pairs of Cr/Au (10/100 nm) electrodes, each of them 9 µm wide and 5 mm long, the gap between two adjacent digits being of 11 µm. In order to reduce the parasitic effects, a passivation coating layer is deposited and patterned in the final fabrication step, being removed from pads and the interdigitated electrically active area (4x5.6 mm$^2$). The entire device was dried in open air, thus achieving a thin film of MoS$_2$ (see Fig. 1).

In our experiments we used a semiconductor characterization system Keithley 4200 SCS and several excitation light sources: (i) a white light lamp with a tunable power and a maximum of 150 W in the visible spectrum consisting of a Xenon arc lamp power supply (Newport, model 69911) and a dedicated Xenon lamp (Newport, model 66902) equipped with an AM1.5G filter, which serves as a sun etalon, (ii) a UV-VIS source in the 215 – 1500 nm spectral range terminated with an optical fiber, consisting of a 43 µW halogen lamp for the VIS spectral domain and a deuterium lamp for UV, with a power of 7 µW, and (iii) a tungsten halogen lamp with a power of 0.5 mW as NIR source, with a spectral domain of 1500 - 2500 nm. The three optical sources cover a very large spectral domain, from 215 nm to 2500 nm. The device in Fig. 1 was positioned on the chuck of the Keithley 4200 SCS and illuminated by different light sources.

First, the *I-V* dependence was collected for the non-illuminated device and the reproducibility was carefully tested; a strong hysteresis was observed, as displayed in Fig. 2. Afterwards, we have illuminated the device with the white light source with a tunable power etalonated in suns and observed that the width of the hysteresis is modulated by light (see Fig. 2), the hysteresis being disabled when the power increases from 1 sun to 5 suns. This modulation is reversible, the hysteresis re-appearing as soon as the power decreases to 1 sun. Finally, in order to investigate the effect of light sources with different spectral ranges, we have directed on the device, separately, the UV and VIS (from the UV-VIS source), as well as the IR radiation source. In this case, as illustrated in Fig. 3, we have observed that the UV and IR radiation sources are able to modulate the width of the hysteretic dependence only in the negative-voltage-part, while the VIS radiation affects both hysteretic branches. The possibility of reversibly modulating the CMS by light is an amazing effect.

3. DISCUSIONS AND CONCLUSIONS

In order to understand the CMS behavior of the thin film consisting of $MoS_2$ monolayers in dark and after illumination with different light sources, some information regarding the structure of such thin films is helpful. Previous structural studies have shown that after drying suspensions of monolayer $MoS_2$, the resulting structure consists of randomly restacked monolayers with a partial rotational order [13], trapping of solvent layers being also possible. The existence of monolayer up to few-layer (due to restacking) $MoS_2$ sheets in the thin film can explain the broad optical response, similar to the results in [14], as well as the formation of quantum dots when regions with a larger number of $MoS_2$ layers and a smaller bandgap are surrounded by regions with fewer layers and larger bandgap. Moreover, $MoS_2$ restacking with partial rotational order leads to the appearance of a large number of defects. In addition, the edges of few-layer $MoS_2$ are known to behave as active sites that favor the adsorbtion of molecules from the environment (which induce *p*-type doping in air [15]), while defect grain boundaries influence the local electronic structure of the material [16]. In addition, the complexity of the system under investigation is enhanced by the Coulomb potential of the shallow trap centers due to adsorbates at the $MoS_2/SiO_2$ interface [15]. All these defects and traps act as recombination centers that lead to drain current hysteresis at forward and reverse scans of the gate voltage in FET configurations [15]. A similar gate hysteresis and *n*-type behavior, maintained even at illumination, has been attributed to the appearance of Schottky barriers in multi-layer $MoS_2$ flakes on $SiO_2$ substrates due to space charge accumulation at Au-$MoS_2$ contacts [17], the interface dipole formation and the generation of gap states due to Au-S interaction leading to Schottky contacts [18].

In agreement with previous results, we propose that the observed hysteresis in our thin film device at dark is caused by charge trapping at defects, adsorbed molecules or grain boundaries. As the drain voltage increases, the trapped electrons become released first and contribute to electrical conduction, the trapped holes being released at a higher voltage. As the holes are released, the current decreases since recombination processes are more probable. Indeed, an asymmetry of electron and hole tunneling from charge traps has been also evidenced in previous works [16,19], and could, in our case, be explained by the different effective masses of electrons and holes in monolayer and few-layer $MoS_2$ [20] and especially by the substantial increase of the hole effective mass in monolayer $MoS_2$ [21]. The identification of charge trapping as the most probable mechanism for generating hysteretic behavior is supported by the fact that no hysteresis is observed (not shown) for either lower or higher concentrations of $MoS_2$ monolayers in the suspension. In the first case electrical conduction is dominated by an incoherent hopping mechanism between mainly non-overlapping $MoS_2$ monolayer sheets, while in the last case a percolative conduction mechanism is likely to occur.

As expected, at illumination the current increases due to photogeneration of electron-hole pairs. Indeed, this fact is observed in Fig. 2 and is consistent with all previous observations. The width of the hysteresis decreases with increasing light power since the height of barriers for holes decrease as the charge in the device due to photogenerated carriers increases. However, the results in Fig. 3, in which the hysteresis width is modified by UV and IR light only for negative voltages suggest that the origin of charge trapping that dominates the conduction at positive and negative voltages is different. In agreement with previous observations [14,18], we associate the charge traps at positive voltages to the gap states at the Au-$MoS_2$ interface, whereas at negative voltages the hysteresis is caused mainly by charge traps due to adsorbed molecules and quantum dots. Indeed, the distribution of charges on MoS$_2$ surfaces due to adsorbates was shown to be drastically affected at illumination. In particular, it was shown that the adsorbed molecules in air, which induce *p*-type doping of MoS$_2$ monolayers in the absence of light illumination, become desorbed from the surface upon illumination in the visible range because the photogenerated holes discharge the adsorbed ions [15]. At UV illumination, the high-energy photons are likely to cause the desorption of a higher number of adsorbed molecules compared to visible photons and thus influence more the current through the device at negative voltages. As for the low-energy IR photons, they are not expected to be absorbed by the high-bandgap mono- and few-layer MoS$_2$ sheets [14], but only by the occasionally restacked portions containing multiple layers. As a consequence the number of photogenerated carriers due to the IR photons is small unless they are absorbed by electrons in the quantum dots in the restacked thin film. This is a possible explanation of why IR illumination only affects the negative-voltage hysteretic branch, the other branch due to mainly deep charge states at the Au-MoS$_2$ interface [18] being almost not influenced by IR illumination.

In conclusion, we have demonstrated that a thin film containing MoS$_2$ monolayers behaves like a CMS. In contrast to previous CMS, the fabrication of this device is simple and does not require homo- or heterostructures. Moreover, the CMSs based on MoS$_2$ display a clear hysteretic behavior, which is strongly modulated by light ranging from UV up to IR. The physical explanation of this unique and unexpected behavior involves charge trapping due to defects, adsorbed molecules or gap states at the contact interfaces, as well as occasional quantum dots. These physical phenomena define the optical response of CMSs based on thin films formed by monolayers of MoS$_2$, response that is different for different spectral ranges and which can modulate the hysteretic behavior of the device.

REFERENCES

[1] Q.H. Wang, K. Kalantar-Zadeh, A. Kis, J.C. Coleman and M. Strano, "Electronics and optical properties of two-dimensional transition metal dichalcogenides," Nature Nanotechnology, 699-712 (2012).

[2] K.F. Mak, C. Lee, J. Hone, J. Shan, and Tony F. Heinz, "Atomically thin $MoS_2$: a new direct-gap semiconductor," Phys. Rev. Lett 105 136805 (2010).

[3] B. Radisavljvic, A. Radenovic, J. Borivio, V. Giacometti, and A. Kis, "Single-layer $MoS_2$ transistor," Nature Nanotechnology 6,147-150 (2011).

[4] O. Lopez-Sanchez, D. Lemke, M. Kayci, A. Radenovic, A. Kis, "Ultrasensitive photodetectors based on monolayer $MoS_2$," Nature Nanotechnology 8, 497-801 (2013).

[5] H.S. Lee, S.-W. Min, Y.-G. Chang, M.K. Park, T. Nam, J.H. Kim, S. Ryn, and S. Inn, "$MoS_2$ nanosheet phototransistors with thickness modulated optical energy gap," Nano Lett. 12, 3695-3700 (2012).

[6] K. Roy, M. Padmanabhan, S. Goswami, T. P. Sai, G. Ramalingam, S. Raghavan and A. Ghosh, "Graphene–$MoS_2$ hybrid structures for multifunctional photoresponsive memory devices," Nature Nanotechnology 8, 826-830 (2013).

[7] K.F. Mak, K. He, J. Shan and Tony F. Heinz, "Control of valley polarization in monolayer $MoS_2$ by optical helicity," Nature Nanotechnology 7, 494-498 (2012).

[8] B. Radisovic, and A. Kis, "Mobility engineering and metal-insulator transitions in monolayer $MoS_2$," Nature Materials, 800-820 (2013).

[9] S. Das, H.-Y. Chen, A. Penimatche, and J. Appenzeller, "High-performance multilayer $MoS_2$ transistor with scandium contacts," Nano Lett. 13, 100-105 (2012).

[10] E. Linn, R. Rosezin, C. Kügeler, and R. Waser, "Complementary resistive switches for passive nanocrossbar memories," Nature Materials 9, 403-406 (2010).

[11] Y. C. Bae, Ah Rahm Lee, J. Bi Lee, J.H. Koo, K.C. Kwon, J. Gun Park, H.S. Im and J.P. Hong, "Oxygen ion drift-induced complementary resistive switching in homo TiO$x$/TiO$y$/TiO$x$ and hetero TiO$x$/TiON/TiO$x$ triple multilayer frameworks," Adv. Funct. Mater. 22, 709-716 (2012).

[12] J. van den Hurk, V. Havel, E. Linn, R. Waser and I. Valov, "Ag/GeSx/Pt-based complementary resistive switches for hybrid CMOS/ nanoelectronic logic and memory architectures," Scientific Reports 3, 1-5 (2013).

[13] P. Joensen, E.D. Crozier, N. Alberding, R.F. Frindt, "A study of single-layer and restacked MoS2 by X-ray diffraction and X-ray absorption spectroscopy", J. Phys. C 20, 4043-4054 (1987).

[14] D.-S. Tsai, K.-K. Liu, D.-H. Lien, M.-L. Tsai, C.-F. Kang, C.-A. Lin, L.-J. Li, J.-H. He, "Few-layer MoS2 with high broadband photogain and fast optical switching for use in harsh environments", ACS Nano 7, 3905-3911 (2013).

[15] W. Zhang, J.-K. Huang, C.-H. Chen, Y.-H. Chang, Y.-J. Cheng, L.-J. Li, "High-gain phototransistors based on a CVD MoS$_2$ monolayer", Adv. Mater. 25, 3456-3461 (2013).

[16] G. Hao, Z. Huang, Y. Liu, X. Qi, L. Ren, X. Peng, L. Yang, X. Wei, J. Zhong, "Electrostatic properties of few-layer MoS$_2$ films", AIP Advances 3, 042125 (2013).

[17] M. Fontana, T. Deppe, A.K. Boyd, M. Rinzan, A.Y. Liu, M. Paranjape, P. Barbara, "Electron-hole transport and photovoltaic effect in gated MoS$_2$ Schottky junctions", Scientific Reports 3, 1634 (2013).

[18] C. Gong, L. Colombo, R.M. Wallace, K. Cho, "The unusual mechanism of partial Fermi level pinning at metal-MoS$_2$ interfaces", Nano Lett. 14, 1714-1720 (2014).

[19] M.S. Choi, G.-H. Lee, Y.-J. Yu, D.-Y. Lee, S.H. Lee, P. Kim, J. Hone, W.J. Yoo, "Controlled charge trapping by molybdenum disulphide and graphene in ultrathin heterostructured memory devices", Nature Commun. 4, 1624 (2013).

[20] F. Zahid, L. Liu, Y. Zhu, J. Wang, H. Guo, "A generic tight-binding model for monolayer, bilayer and bulk MoS$_2$", AIP Advances 3, 052111 (2013).

[21] W. Jin, P.-C. Yeh, N. Zaki, D. Zhang, J.T. Sadowski, A. Al-Mahboob, A.M. van der Zande, D.A. Chenet, J.I. Dadap, I.P. Herman, P. Sutter, J. Hone, R.M. Osgood, Jr., "Direct measurement of the thickness-dependent electronic band structure of MoS$_2$ using angle-resolved photoemission spectroscopy", Phys. Rev. Lett. 111, 106801 (2013).

FIGURE CAPTIONS

Fig. 1 (a) Microscope photo of the device before (left) and after (right) MoS$_2$ deposition. (b) Magnified microscope photo of the device before (left) and after (right) MoS$_2$ deposition in green light.

Fig. 2 The *I-V* dependence in white light at various optical powers.

Fig. 3 The *I-V* dependence for UV, VIS and IR illuminations.

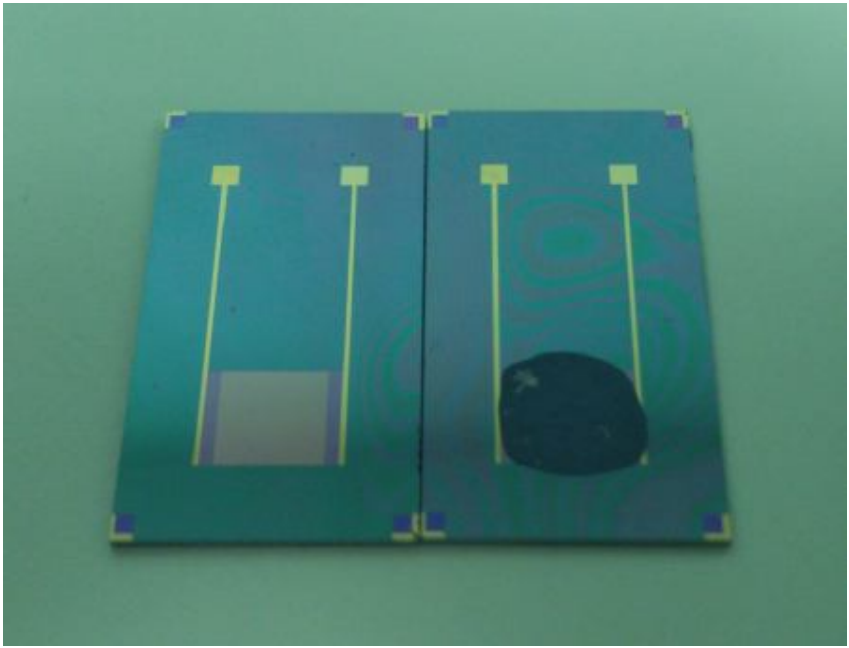

(a)

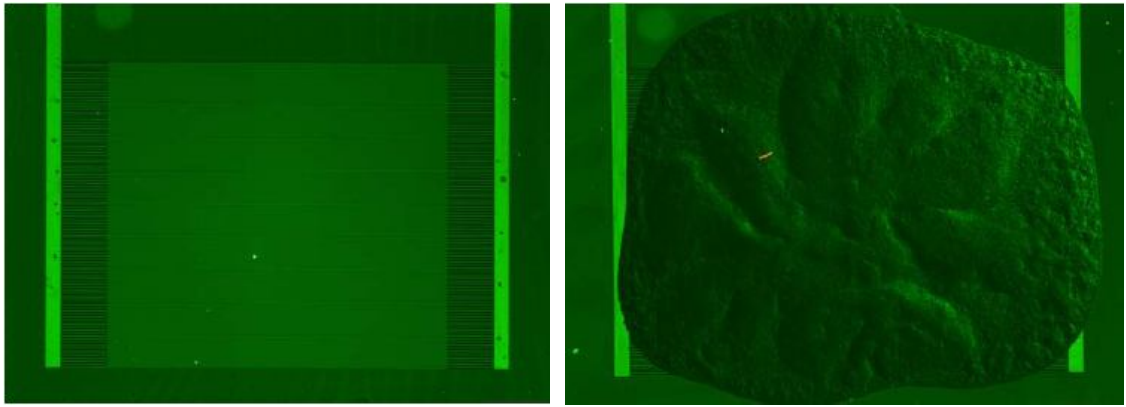

(b)

Fig. 1

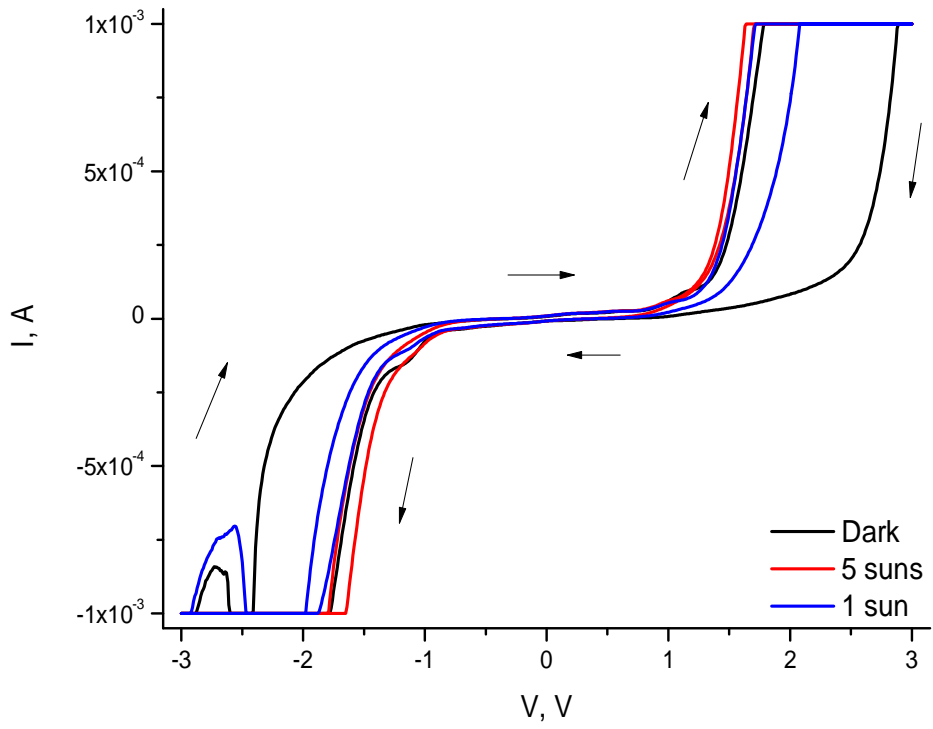

Fig. 2

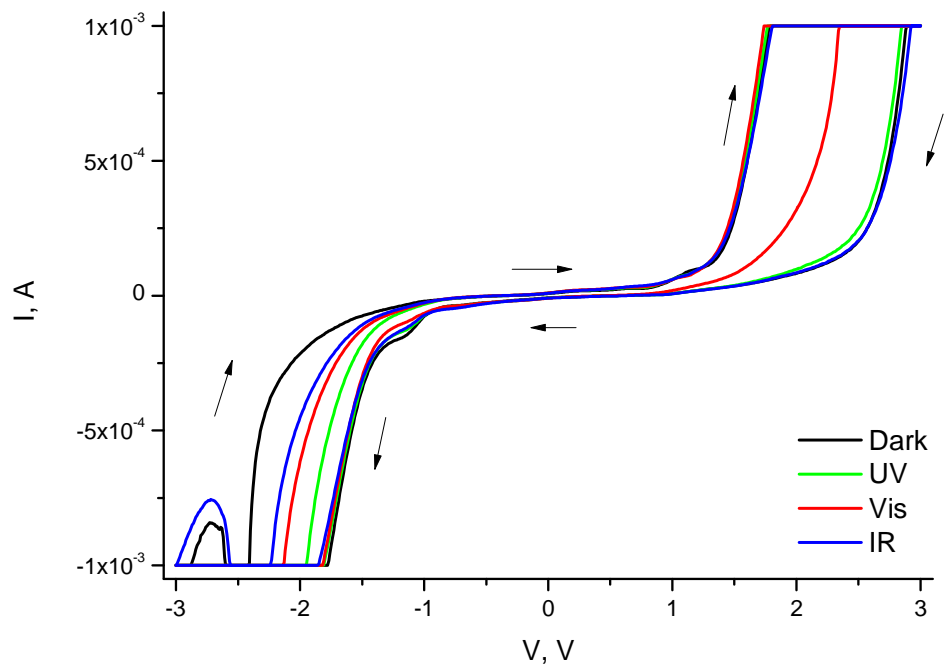

Fig. 3